\documentclass[twocolumn,prd0000000000000000,superscriptaddress,altaffilletter,nofootinbib]{revtex4-1}
\usepackage{amsmath}
\usepackage{amssymb}
\usepackage{amsfonts}
\usepackage{comment}
\usepackage{graphicx, caption}
\usepackage{hyperref}
\usepackage{enumerate}
\usepackage{ulem}
\usepackage[nolist]{acronym}
\usepackage[usenames,dvipsnames]{xcolor}
\usepackage{xspace}
\usepackage{tikz}
\usepackage{ulem}
\usepackage{siunitx}
\DeclareSIUnit \parsec {pc}
\allowdisplaybreaks
\usepackage{multirow}
\usepackage{bm}
\usepackage{romannum}
\usepackage{mathtools}
\newcommand{\tabref}[1]{Tab. \ref{#1}}

\newcommand{\secref}[1]{Sec.~\ref{#1}}

\usepackage{soul}
\usepackage{xcolor}
\usepackage{caption}
\newcommand{\Msun}{M_{\odot}}

\newcommand{\flow}{f_{\mathrm{low}}}
\newcommand{\fhigh}{f_{\mathrm{high}}}
\newcommand{\ceil}[1]{\left\lceil #1 \right\rceil}
\newcommand{\floor}[1]{\left\lfloor #1 \right\rfloor}

\usetikzlibrary{arrows,shapes,trees,decorations.pathreplacing}
\begin{document}
\pagenumbering{arabic}
\title{Accelerating gravitational-wave parameterized tests of General Relativity \\
using a multiband decomposition of likelihood} 

\author{Naresh Adhikari}
\email{naresh@uwm.edu}
  \affiliation{Department of Physics, University of Wisconsin-Milwaukee, Milwaukee, WI 53201, USA}
\author{Soichiro Morisaki}
\email{morisaki@uwm.edu}
  \affiliation{Department of Physics, University of Wisconsin-Milwaukee, Milwaukee, WI 53201, USA}
\date{\today}

\begin{abstract}
The detection of gravitational waves from compact binary coalescence (CBC) has allowed us to probe the strong-field dynamics of General Relativity (GR).
Among various tests performed by the LIGO--Virgo--KAGRA collaboration are parameterized tests, where parameterized modifications to GR waveforms are introduced and constrained.
This analysis typically requires the generation of more than millions of computationally expensive waveforms.
The computational cost is higher for a longer signal, and current analyses take weeks--years to complete for a binary neutron star (BNS) signal.
In this work, we present a technique to accelerate the parameterized tests using a multiband decomposition of likelihood, which was originally proposed to accelerate parameter estimation analyses of CBC signals assuming GR by one of the authors.
We show that our technique speeds up the parameterized tests of a 1.4$\Msun$–1.4$\Msun$ BNS signal by a factor of
$\mathcal{O}(10)$ for a low-frequency cutoff of $20$ Hz.
We also verify the accuracy of our method using simulated signals and real data.
\end{abstract}

\maketitle

\section{Introduction} \label{sec:introduction}

The gravitational-wave era started with the discovery of the gravitational wave signal from the binary black hole merger, GW150914 \cite{LIGOScientific:2016aoc}, by the advanced LIGO detectors \cite{LIGOScientific:2014pky, aLIGO:2020wna, Tse:2019wcy}. The binary neutron star (BNS) signal, GW170817 \cite{LIGOScientific:2017vwq}, was observed two years later by the advanced LIGO and advanced Virgo \cite{VIRGO:2014yos, Virgo:2019juy} detectors. It became the first example of multimessenger observations involving gravitational waves \cite{LIGOScientific:2017ync,LIGOScientific:2017zic}. Recently, the first-ever observations of mergers of two distinct compact objects, i.e., a neutron star and a black hole, were also achieved \cite{LIGOScientific:2021qlt}, completing the search for a gravitational-wave signal originating from all three distinct classes of compact mergers.
The detected CBC signals enabled us to test General Relativity (GR) in the strong-field regime. 
Various tests have been proposed and applied to the detected signals by the LIGO--Virgo--KAGRA collaboration (LVK) \cite{LIGOScientific:2016lio, LIGOScientific:2018dkp, LIGOScientific:2019fpa, LIGOScientific:2020tif, LIGOScientific:2021sio} and others \cite{Isi:2019aib, Takeda:2020tjj, Shoom:2021mdj}.

Among various tests performed by LVK are the parameterized tests \cite{Arun:2006hn, Li:2011cg, Mishra:2010tp, Agathos:2013upa}. In the parameterized tests, parameterized non-GR modifications are introduced to GR waveforms, and the parameters governing the modifications are constrained. The non-GR parameters consist of inspiral parameters and post-inspiral parameters. The inspiral parameters parameterize relative or absolute shifts of the inspiral post-Newtonian coefficients, while the post-inspiral parameters parameterize relative shifts of the post-inspiral phenomenological parameters. For informative constraints to be obtained efficiently, typically only one of those non-GR parameters is allowed to deviate and constrained in a single analysis \cite{Sampson:2013lpa, LIGOScientific:2016lio}. Such a single-parameter test is known to be robust to ignorance of higher-order corrections  \cite{Perkins:2022fhr}. Recent works \cite{Shoom:2021mdj, Saleem:2021nsb} also showed that multiple parameters can be investigated simultaneously using principal component analysis. Those parameterized modifications can incorporate modifications predicted by various alternative theories of gravity, and we can map those constraints to filter such non-GR theories as a post-processing step \cite{Yunes:2016jcc}. 

The parameterized tests typically employ stochastic sampling and require more than millions of likelihood evaluations.
Each likelihood evaluation requires the evaluations of waveform values at all the frequency points considered, which is the dominant cost.
Since the frequency points are sampled with an interval of $1/T$, where $T$ is the duration of data, more waveform evaluations are required for a longer signal.
For a $1.4\Msun$–$1.4\Msun$ BNS signal, current analyses take weeks--years without any approximate methods.
This is going to be a serious problem when the sensitivities of detectors are improved and BNS signals are detected more frequently.
The same problem arises for parameter estimation analyses of CBCs assuming GR, and various techniques have been proposed to reduce the computational cost of waveform generation \cite{Canizares:2014fya, Smith:2016qas, Morisaki:2020oqk, Vinciguerra:2017ngf, Morisaki:2021ngj, Zackay:2018qdy, Cornish:2021lje}.

Among the various rapid parameter estimation techniques, a recent work considers a multiband decomposition of the gravitational-wave likelihood \cite{Morisaki:2021ngj}, which exploits the chirping nature of the CBC signals and speeds up the parameter estimation of a BNS signal by more than an order of magnitude.
Since the signal frequency increases with time, the time to merger, $\tau(f)$, decreases with frequency $f$.
This implies that the likelihood can be approximated into a form that can be computed with waveform values at frequency points sampled with a variable interval proportional to $1 / \tau(f)$.
This approximation drastically reduces the number of waveform evaluations at high frequency.
A similar idea has been utilized for speeding up the matched-filter analysis for detection of CBC signals \cite{marion:in2p3-00014163, Buskulic:2010zz, Cannon:2011vi}.

In this paper, we apply the multiband decomposition technique to parameterized tests of GR.
In \secref{sec:basics}, we briefly explain parameterized tests of GR and the multiband decomposition method for a GR signal. 
To extend the multiband decomposition method to parameterized tests, we need to consider modifications of $\tau(f)$ caused by non-GR modifications in waveforms.
In \secref{sec:methods}, we derive the modified $\tau(f)$ and investigate the speed-up gains of our technique for a BNS signal.
In \secref{sec:validation} we study the accuracy of our technique using simulated BNS signals and real data. 
Finally, we conclude our work in \secref{sec:conclusion}.

\section{Basics} \label{sec:basics}

In this section, we explain the parameterized tests of GR and the multiband decomposition technique for rapid parameter estimation.

\subsection{Parameterized tests of GR}

The dominant quadrupole moment of gravitational waves is of the following form in the frequency domain,
\begin{equation}
\label{strain}
\tilde{h}(f) = A(f)e^{i{\Phi(f)}},
\end{equation}
where $A(f)$ and $\Phi(f)$ denote signal amplitude and phase respectively. The phase evolution of the early-inspiral part is calculated via the Post-Newtonian (PN) expansion \cite{Will:2011nz, Blanchet:2013haa}, which is an expansion with respect to a small orbital velocity $v/c$.
A term with the order of $\mathcal{O}((v/c)^n)$ relative to the leading order is referred to as $(n/2)$PN. In GR, the phase up to the 3.5PN order is given by \cite{Buonanno:2009zt, Sathyaprakash:1991mt, Blanchet:1994ez},
\begin{equation}
\label{phase}
\begin{aligned}
&\Phi^{\mathrm{GR}}(f) = \\
&2\pi f t_{c} - \phi_{c} - \frac{\pi}{4} + \sum_{j = 0}^7 \left[ \varphi^{\mathrm{GR}}_{j} + \varphi_{j}^{\mathrm{GR}(l)} \ln{f} \right] f^{(j-5)/3},
\end{aligned}
\end{equation}
where $t_c$ and $\phi_{c}$ denote the coalescence time and phase respectively, and $\varphi^{\mathrm{GR}}_{j}$ and $\varphi_{j}^{\mathrm{GR}(l)}$ are $(j/2)$PN coefficients depending on component masses $m_{1}$, $m_{2}$ and spins $\vec{S_{1}}$, $\vec{S_{2}}$.
$\varphi^{\mathrm{GR}}_{j}$ is vanishing for $j=1$ and $\varphi^{\mathrm{GR}(l)}_j$ is vanishing except for $j=5, 6$.

In the parameterized tests, parameterized deformations of non-zero PN coefficients are introduced by \cite{Arun:2006hn, Li:2011cg, Mishra:2010tp, Agathos:2013upa},
\begin{equation*}
\varphi^{\mathrm{GR}}_{j} \to \left[ 1 + \delta \hat \varphi_{j}\right]\varphi^{\mathrm{GR}}_{j}, ~~~\varphi^{\mathrm{GR}(l)}_{j} \to \left[ 1 + \delta \hat \varphi^{(l)}_{j}\right] \varphi^{\mathrm{GR}(l)}_{j},
\end{equation*}
where $\delta \hat \varphi_{j}$ and $\delta \hat \varphi^{(l)}_{j}$ are non-GR parameters quantifying relative shifts of GR inspiral phasing.
In addition to the relative shifts, absolute shifts are introduced to the $-1$PN,
\begin{equation}
\varphi_{-2} f^{-7/3} = \frac{3 \delta \hat \varphi_{-2}}{128} \eta^{2/5} \left(\frac{\pi G \mathcal{M} f}{c^3} \right)^{-7/3},
\end{equation}
and $0.5$PN,
\begin{equation}
\varphi_{1} f^{-4/3} = \frac{3 \delta \hat \varphi_{1}}{128 \eta^{1/5}} \left(\frac{\pi G \mathcal{M} f}{c^3} \right)^{-4/3},
\end{equation}
where $\mathcal{M}$ and $\eta$ are chirp mass and symmetric mass ratio respectively,
\begin{equation}
\mathcal{M} = \frac{(m_1 m_2)^{3/5}}{(m_1 + m_2)^{1/5}},~~~~~\eta = \frac{m_1 m_2}{(m_1 + m_2)^2}.
\end{equation}
The $-1$PN term is to model gravitational dipole radiation predicted by alternative theories of gravity \cite{eardley1975observable}.
The full list of inspiral non-GR parameters is,
\begin{equation*}
\{  \delta \hat \varphi_{-2},  \delta \hat \varphi_{0},   \delta \hat \varphi_{1},   \delta \hat \varphi_{2},   \delta \hat \varphi_{3},   \delta \hat \varphi_{4},  \delta \hat \varphi_{5}^{(l)},   \delta \hat \varphi_{6},\delta \hat \varphi_{6}^{(l)},   \delta \hat \varphi_{7}\}.
\end{equation*}
The 2.5PN parameter $\delta \hat \varphi_{5}$ is not included since it is completely degenerate with $\phi_{c}$.

In addition to the deformations of inspiral phase, parameterized deformations of post-inspiral phase are also considered.
The IMRPhenom waveform model \cite{Ajith:2007kx, Husa:2015iqa, Khan:2015jqa} employs phase ansatz parameterized by $\beta_i ~(i=0,1,2,3)$ for intermediate stage and by $\alpha_i~(i=0,1,2,3,4)$ for merger-ringdown stage.
Relative shifts to those parameters are introduced in a similar way, which are parameterized by $\delta \hat \beta_i$ and $\delta \hat \alpha_i$.
The full list of post-inspiral non-GR parameters as considered in \cite{LIGOScientific:2016lio, LIGOScientific:2019fpa, LIGOScientific:2020tif, LIGOScientific:2021sio} is,
\begin{equation*}
\{\delta \hat \alpha_2, \delta \hat \alpha_3, \delta \hat \alpha_4, \delta \hat \beta_2, \delta \hat \beta_3\}.
\end{equation*}
For meaningful constraints to be obtained efficiently, typically only one of those 15 non-GR parameter is allowed to deviate and constrained in a single analysis.
Parameterized deformations of amplitude can also be considered, but they are difficult to be measured with the current generation of detectors \cite{VanDenBroeck:2006qu, VanDenBroeck:2006ar, OShaughnessy:2013zfw}.
                 
The non-GR parameters are estimated or constrained via Bayesian inference.
In the Bayesian inference, posterior distribution $p\left(\bm{\theta}|\{\bm{d}_i\}\right)$ is calculated via the Bayes theorem:
\begin{equation}
p\left(\bm{\theta}|\{\bm{d}_i\}\right) \propto  \mathcal{L}\left(\{\bm{d}_i\} |\bm{\theta}\right)\pi(\bm{\theta}),
\end{equation}
where $\bm{d}_i$ denotes the data taken from the $i$--th detector, $\bm{\theta}$ the set of model parameters consisting of one of the non-GR parameters and GR parameters, $\pi(\bm{\theta})$ the prior distribution function determined from our belief or prior knowledge on $\bm{\theta}$, and $\mathcal{L}(\bm{d}|\bm{\theta})$ the likelihood function. For the likelihood, the Gaussian-noise likelihood function is typically used \cite{Thrane:2018qnx, Christensen:2022bxb},
\begin{equation}
\mathcal{L}(\{\bm{d}_i\}|\bm{\theta}) \propto \exp\left[ - \frac 1 2 \sum_{i} \|\bm{d}_{i} - \bm{h}_{i}( \bm{\theta}) \|^2_i \right], \label{eq:loglikelihodd}
\end{equation}
\noindent where $\bm{h}_i$ is a model signal observed at the $i$--th detector. $\| \cdot \|^2 = \left( \cdot, \cdot \right)$ is the norm induced by the inner product,
\begin{equation}
\left ( \bm{a}, \bm{b} \right )_i = \frac{4}{T} \Re \left[\sum_{k=\flow T}^{\fhigh T} \frac{\tilde{a}^\ast(f_{k}) \tilde{b}(f_{k})}{S_i(f_{k})}\right],
\label{eq:inner_prod}
\end{equation}
where $\flow$ and $\fhigh$ are the low-- and high--frequency cutoffs of the analysis respectively, $T$ is the duration of data, $S_i(f)$ is the noise power spectral density of the $i$--th detector, and $f_k\equiv k/T$ is the $k$--th frequency bin. The logarithm of likelihood can be written as
\begin{equation}
\begin{aligned}
 & \ln \mathcal{L}(\bm{d}|\bm{\theta})  =  \sum_i \left[\left( \bm{d}_i, \bm{h}_i( \bm{\theta}) \right)_i - \frac 1 2 \left( \bm{h}_i( \bm{\theta}), \bm{h}_i(\bm{\theta}) \right)_i \right] \\
 & ~~~~~~~~~~~~~~~ + \text{const.},
 \label{eq:nonconstantll}
\end{aligned}
\end{equation}                      
where the constant part does not depend on $\bm{\theta}$ and is irrelevant for stochastic sampling.

The inference is typically done via stochastic sampling methods, such as Markov-Chain Monte Carlo (MCMC) \cite{Metropolis:1953am, Hastings:1970aa} and nested sampling \cite{skilling2006}.
The non-constant term is computed millions of times during stochastic sampling. 
As evident from Eqs. \eqref{eq:nonconstantll} and \eqref{eq:inner_prod}, each likelihood evaluation requires evaluations of waveform values at all the frequency points from $\flow$ to $\fhigh$.
Those waveform evaluations are typically the dominant cost of analysis.
The cost is proportional to the number of frequency points,
\begin{equation}
K_{\mathrm{orig}} = (\fhigh - \flow) T + 1,
\label{eq:korig}
\end{equation}
and higher for a longer signal.

\subsection{Multiband decomposition} \label{sec:multiband}

In the multiband decomposition method, the total frequency range is divided into $B$ overlapping frequency bands $f^{(b)}_{\mathrm{s}} \leq f \leq f^{(b)}_{\mathrm{e}}~~~(b=0,1,\dots,B-1)$.
The start and end frequencies are determined based on a user-specified sequence of durations, $T=T^{(0)}>T^{(1)}>\dots>T^{(B-1)}$.
First, the following equation is solved with respect to $f^{(b)}$ for each $b \in \{1,2,\dots,B-1\}$,
\begin{equation}
\tau(f^{(b)}) + L \sqrt{-\tau'(f^{(b)})} = T^{(b)} + t_{\mathrm{c, min}} - T, \label{eq:band_equation}
\end{equation}                                                                  
where $\tau(f)$ is a reference time-to-merger from a gravitational wave frequency $f$.
$L$ is a user-specified constant controlling the accuracy of the approximation.
A larger value of $L$ gives more accurate likelihood values.
$t_{\mathrm{c, min}}$ is the minimum coalescence time in the prior range. $L=5$ and $T-t_{\mathrm{c, min}}=2.12\,\mathrm{s}$ are used throughout this paper, following \cite{Morisaki:2021ngj}.
The start and end frequencies are determined as
\begin{align}
&f^{(b)}_{\mathrm{s}} = \begin{cases}
\displaystyle f_{\mathrm{low}}, & (b=0) \\
\displaystyle f^{(b)} - \frac{1}{\sqrt{\tau' (f^{(b)})}}, & (b > 0)
\end{cases} \\
&f^{(b)}_{\mathrm{e}} = \begin{cases}
f^{(b + 1)}, & (b < B-1) \\
f_{\mathrm{high}} + \Delta f_{\mathrm{high}}. & (b = B - 1)
\end{cases}
\end{align}
This way of constructing frequency bands guarantees that the inverse Fourier transform of $\tilde{h}(f)$ starting from $f^{(b)}_{\mathrm{s}}$ has a duration shorter than $T^{(b)}$.
$\Delta f_{\mathrm{high}} > 0$ is required to avoid the loss of accuracy caused by the abrupt termination of a waveform.

With the frequency bands constructed, $(\bm{d}_i, \bm{h}_i)_i$ is approximated into the following form,
\begin{equation}
\begin{aligned}
&(\bm{d}_i, \bm{h}_i(\bm{\theta}))_i \simeq  \\
&\sum^{B-1}_{b=0} \frac{4}{T^{(b)}} \Re\left[\sum_{k=\ceil{f^{(b)}_{\mathrm{s}} T^{(b)}}}^{\floor{f^{(b)}_{\mathrm{e}} T^{(b)}}} w^{(b)} (f^{(b)}_k) \tilde{D}^{(b)\ast}_{i,k} \tilde{h}_i(f^{(b)}_k;\bm{\theta}) \right],
\end{aligned}
\end{equation}
where $w^{(b)}(f)$ is a smooth window function extracting waveform values in the $b$--th frequency band, $\tilde{D}^{(b)}_{i,k}$ is a quantity calculated from data and power spectral density, and
\begin{equation}\label{fT}           
f^{(b)}_k \equiv \frac{k}{T^{(b)}}.
\end{equation}
The sum over a high-frequency band requires waveform values only at downsampled frequencies whose interval is $1/T^{(b)}$, and hence fewer waveform evaluations.
The number of waveform evaluations required for a single evaluation of $(\bm{d}_i, \bm{h}_i(\bm{\theta}))_i$ is reduced to
\begin{equation}
K_{\mathrm{MB}} = \sum_{b=0}^{B - 1} \left( \floor{f^{(b)}_{\mathrm{e}} T^{(b)}} - \ceil{f^{(b)}_{\mathrm{s}} T^{(b)}} + 1 \right).
\end{equation}
     
There were two approximate methods proposed to compute $(\bm{h}_i(\bm{\theta}), \bm{h}_i(\bm{\theta}))_i$ with fewer waveform evaluations.
One method is referred to as {\it Linear Interpolation}, which approximates $|\tilde{h}_i(f;\bm{\theta})|^2$ as a linear interpolation of the squares of downsampled waveform values. This works well if the waveform model contains only dominant quadrupole moments, where $|\tilde{h}_i(f;\bm{\theta})|^2$ is a smooth function.
The other method is referred to as {\it IFFT-FFT}, which works even if the waveform model contains multiple moments.
In either case, $(\bm{h}_i(\bm{\theta}), \bm{h}_i(\bm{\theta}))_i$ is computed with waveform values at the $K_{\mathrm{MB}}$ frequency points, and no additional waveform evaluations are required.
Thus, the cost of a single likelihood evaluation is reduced by $K_{\mathrm{orig}} / K_{\mathrm{MB}}$.

\section{Extension to parameterized tests} \label{sec:methods}

In the previous work \cite{Morisaki:2021ngj}, which applies the multiband decomposition method to the analysis of a GR signal, the 0PN formula of $\tau(f)$ in GR is used for solving \eqref{eq:band_equation}.
For extending the previous work to parameterized tests of GR, we need to take into account corrections of $\tau(f)$ from the parameterized modifications of inspiral phasing.
In this section, we derive the modified formula of $\tau(f)$ taking them into account.
We also apply it for setting up frequency bands and study the speed-up gains of our method for a typical BNS signal.

\subsection{Modified time to merger} \label{sec:timetomerger}

In order to get the modified time-to-merger formula, we use the following condition in accord to the stationary phase approximation \cite{Sathyaprakash:1991mt, Poisson:1995ef, Creighton:2011zz, Maggiore:2007ulw}:
\begin{equation}
    \Phi(f) = - \Psi(t (f)) + 2 \pi f t (f) + \frac{\pi}{4},
\end{equation}
where $\Psi(t)$ is the phase of a time-domain waveform and $t(f)$ is the time at which $\Psi'(t) = 2 \pi f$. 
Hence, we can now relate $t(f)$ to the derivative of $\Phi(f)$,
\begin{equation}
t(f) = \frac{\Phi'(f)}{2 \pi}.
\end{equation}
With the phase formula at inspiral part, we obtain the following modified time-to-merger formula,
\begin{align}
\tau(f) &= t_{c} - t(f)  \nonumber \\
&=\frac{1}{2 \pi} \Bigg \{ \frac { 7 \varphi_{-2} } {3 f^{10/3}} + \frac{ 5 (1 + \delta \hat \varphi_{0})\varphi_{0}^{\mathrm{GR}} }{3f^{8/3}} + 
\frac { 4 \varphi_{1} } {3 f^{7/3}} \nonumber \\
&+ \frac{(1 + \delta \hat \varphi_{2}) 
 \varphi_{2}^{\mathrm{GR}}}{f^{2}} + \frac{ 2(1 + \delta \hat \varphi_{3}) \varphi_{3}^{\mathrm{GR}}}{3f^{5/3}}  + \frac{ (1 + \delta \hat \varphi_{4})\varphi_{4}^{\mathrm{GR}}}{3f^{4/3}} \nonumber \\
&- \frac{ (1 + \delta \hat \varphi_{5}^{(l)})\varphi_{5}^{(l){\mathrm{GR}}}}{f} \nonumber \\
&- \frac{(1 + \delta \hat \varphi_{6}) \varphi_{6}^{\mathrm{GR}} + (1 + \delta \hat \varphi_{6}^{(l)}) \varphi_{6}^{(l){\mathrm{GR}}} (\ln f + 3)}{3f^{2/3}} \nonumber \\
&- \frac{ 2 (1 + \delta \hat \varphi_{7}) \varphi_{7}^{\mathrm{GR}}}{3f^{1/3}} \Bigg \}. \label{tomerge}
\end{align}
Since the time-to-merger is predominantly determined by the terms up to 0PN, we ignore terms higher than that order, and employ the following formula,
\begin{align}
\tau(f) &= \frac { 7 \varphi_{-2} } {6 \pi f^{10/3}} + \frac{ 5 (1 + \delta \hat \varphi_{0})\varphi_{0}^{\mathrm{GR}} }{6 \pi f^{8/3}} \\
&= \frac{7 \delta \hat \varphi_{-2}}{256 \pi} \eta^{2/5} \left(\frac{\pi G  \mathcal{M}}{c^3}\right)^{{-7}/{3}} f^{{-10}/{3}} \nonumber \\
&~~~~+ \frac{5(1 + \delta \hat \varphi_{0})}{256 \pi} \left(\frac{\pi G  \mathcal{M}}{c^3}\right)^{{-5}/{3}} f^{{-8}/{3}}. \label{finaltau}
\end{align}
If higher-order multiple moments are present, the same formula with the frequency rescaling, $f \to 2 f / m$, is used, where $m$ is the maximum magnetic number of the moments.

\begin{figure*}
                 \includegraphics[width = \columnwidth]{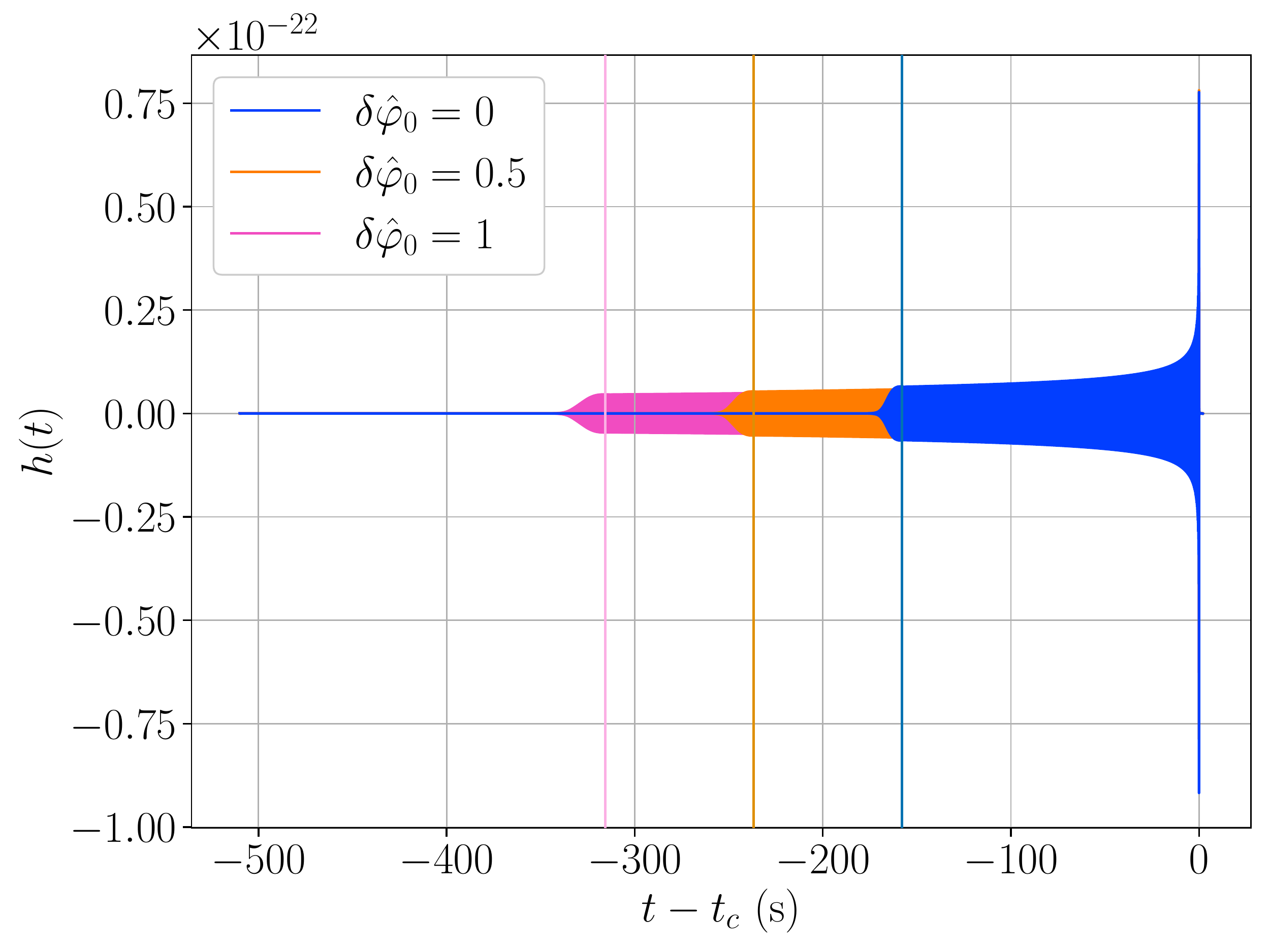}
                 \includegraphics[width = \columnwidth]{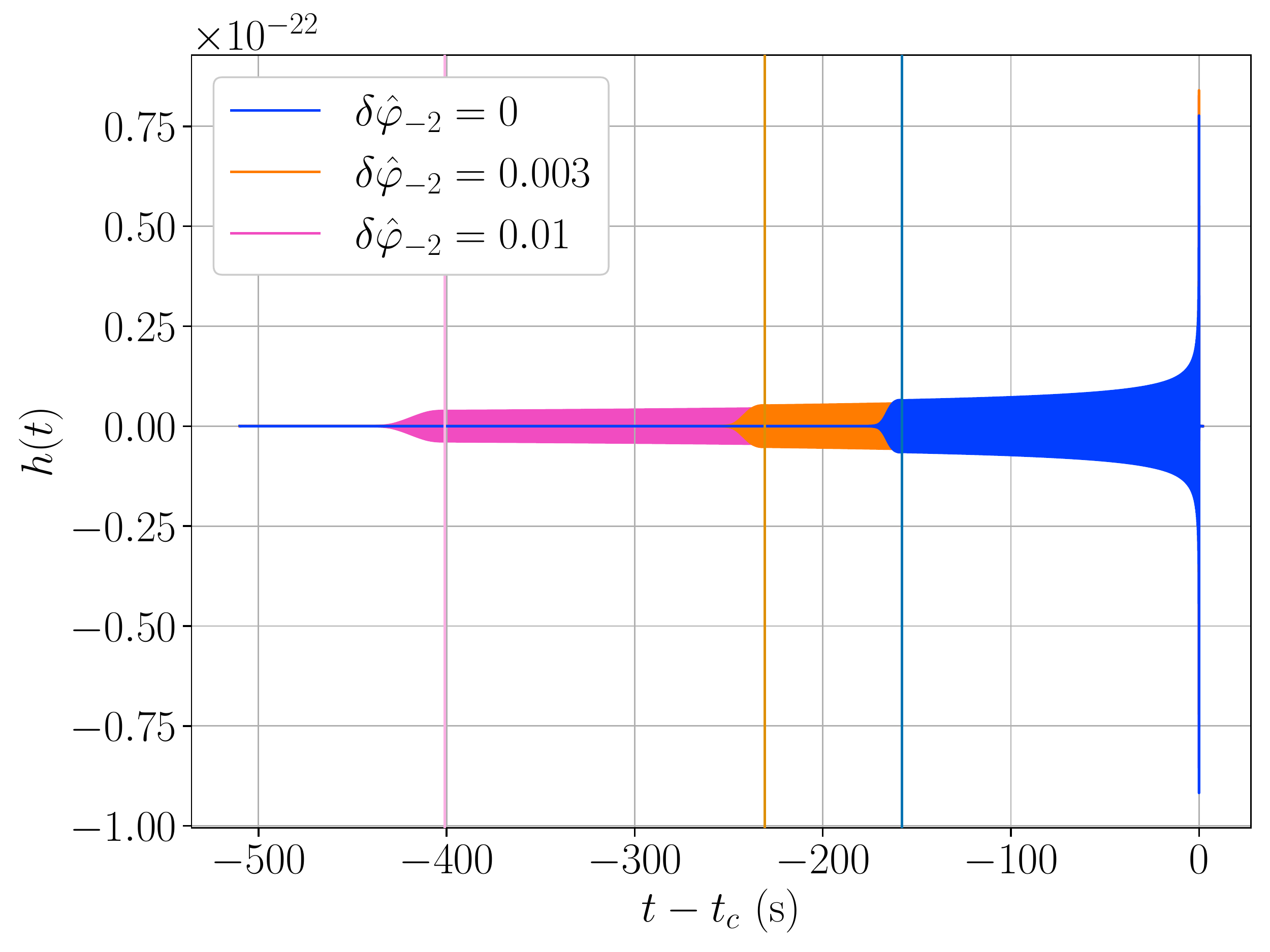}
                 \captionsetup{labelfont=bf,
              justification=raggedright,
              singlelinecheck=false}
                 \caption{Time-domain gravitational waveforms of non-spinning $1.4\Msun$--$1.4\Msun$ BNS starting from $20\,\si{\hertz}$ with various values of $\delta \hat{\varphi}_0$ (left) or $\delta \hat{\varphi}_{-2}$ (right). Vertical lines represent durations calculated by the up-to-0PN time-to-merger formula \eqref{finaltau}.}
                 \label{fig:tmerge}
 \end{figure*}

For validating our approximate time-to-merger formula, we numerically calculate time-domain waveforms incorporating higher-order PN terms and compare their durations with predictions from our formula.
Figure \ref{fig:tmerge} shows time-domain waveforms for non-spinning $1.4\Msun$--$1.4\Msun$ BNS with various values of $\delta \hat{\varphi}_0$ or $\delta \hat{\varphi}_{-2}$.
They are calculated as the inverse Fourier transforms of frequency-domain waveforms from $20\,\si{\hertz}$ to $1024\,\si{\hertz}$ that include terms up to the 3.5PN order in phase and the leading-order term in amplitude.
The GR phase coefficients have been calculated with \texttt{SimInspiralTaylorF2AlignedPhasing} implemented in the LIGO Algorithmic Library (LAL) \cite{lalsuite}.
The vertical lines represent predictions from our approximate time-to-merger formula with $f=20\,\si{\hertz}$. 
As seen in the figure, vertical lines accurately locate the time when waveforms start, demonstrating that our approximate time-to-merger is accurate enough.

Evaluating Eq. \eqref{finaltau} demands a choice on the values of $\mathcal{M}$, $\eta$, $\delta \hat \varphi_{0}$, and $\delta \hat \varphi_{-2}$.
To guarantee that the duration of each frequency band is long enough for any template waveform generated during stochastic sampling, their values are chosen to maximize $\tau(f)$.
Hence, the minimum value of $\mathcal{M}$ and the maximum values of $\eta$, $\delta \hat \varphi_{0}$, and $\delta \hat \varphi_{-2}$ within the explored parameter space are chosen.
The maximum value of $\eta$ is typically $1/4$, which corresponds to $m_1=m_2$.

From Eq. \eqref{finaltau}, it is clear that $\tau(f)$ becomes negative for $\delta \hat \varphi_{0} < -1$ or $\delta \varphi_{-2} < 0$ unless the other terms are significant enough to compensate it.
In this case, the template waveform is an inverse-chirp waveform, which starts from $t=t_{\mathrm{c}}$ and whose frequency simply decreases.
The multiband approximation clearly breaks down for this type of waveform since it assumes that the signal frequency simply increases.
Even without the multiband approximation, the inverse-chirp signal is not properly analyzed in  analysis with the standard data conditioning \cite{veitch:2014wba}, where only data up to $\sim 2$ seconds after $t_{\mathrm{c}}$ are analyzed.

Since the higher-order terms are ignored in Eq. \eqref{finaltau}, the huge deviations from GR in one or more of the higher-order terms can make the approximate time-to-merger formula inaccurate. In a typical analysis, the explored range of $\mathcal{M}$ is much wider than the width of its marginal posterior distribution, and the time to merger computed with the minimum $\mathcal{M}$ in the explored range is large enough to construct conservative frequency bands. The same argument can be made for hidden modifications with non-PN frequency dependences considered in \cite{Li:2011cg}. Also, it is straightforward to take into account higher-order terms from Eq. \ref{tomerge} when huge deviations of higher-order terms are considered.

\subsection{Speed-up gains}\label{sec:speedup}

Table \ref{tab:speed_up} shows speed-up gains of our multiband technique for a $1.4\Msun$--$1.4\Msun$ BNS signal with several choices of $T$ and $\delta \hat{\varphi}_i$ used for calculating $\tau(f)$.
For each case in the table, frequency bands were set up with the algorithm described in \secref{sec:multiband} and Eq. \eqref{finaltau} calculated with $m_1=m_2=1.4\Msun$ and $\delta \hat{\varphi}_i$ of the row.
The total frequency range is $20$--$2048\,\si{\hertz}$, and the durations of bands are powers of two, $\{T^{(b)}\}_{b=0}^{B-1} = \{T,~T/2,~T/4,~\cdots,~4\,\si{\second}\}$.
The speed-up gain is estimated by the reduction of frequency points, $K_{\mathrm{orig}} / K_{\mathrm{MB}}$.
For setting up frequency bands, we utilized the existing implementation of the multiband decomposition method, \texttt{MBGravitationalWaveTransient}, available in the \texttt{bilby} \cite{Ashton:2018jfp, Romero-Shaw:2020owr} software.

For this study, we consider the 3 choices of $\delta \hat{\varphi}_i$: GR ($\delta \hat{\varphi}_0=\delta \hat{\varphi}_{-2}=0$) for reference, $0\mathrm{PN}$ ($\delta \hat{\varphi}_0=20,~\delta \hat{\varphi}_{-2}=0$), and $-1\mathrm{PN}$ ($\delta \hat{\varphi}_0=0,~\delta \hat{\varphi}_{-2}=1$).
$\delta \hat{\varphi}_0=20$ or $\delta \hat{\varphi}_{-2}=1$ is the maximum of its range explored by LVK analyses, which we have found in configuration files available at \cite{tgrsamples}.
In a standard LVK parameterized test, the duration of analyzed data is the same as that used for GR parameter estimation regardless of the explored range of a non-GR parameter.
For a $1.4\Msun$--$1.4\Msun$ BNS signal, $T=256\,\si{\second}$.
In either case with $T=256\,\si{\second}$ in the table, the speed-up gain is $\mathcal{O}(10)$.
The speed-up gain for $0\mathrm{PN}$ or $-1\mathrm{PN}$ is smaller than that for GR because $\tau(f)$ gets larger due to the non-GR modification.

To properly analyze any waveform within the explored range of a non-GR parameter, the data duration should be longer than the longest duration of the waveform with the allowed non-GR modifications.
If data durations are determined in that conservative way, $T=4096\,\si{\second}$ and $T=32768\,\si{\second}$ for $0\mathrm{PN}$ and $-1\mathrm{PN}$ respectively.
With that conservative choice of $T$, the speed-up gain gets larger and is $\mathcal{O}(10^2)$ for either case.

\begin{table*}[t]
\setlength{\tabcolsep}{12pt}
\centering
\captionsetup{labelfont=bf,
              justification=raggedright,
              singlelinecheck=false}
\caption{The numbers of original frequency points $K_{\mathrm{orig}}$, the numbers of multibanded frequency points $K_{\mathrm{MB}}$, and speed-up gains $K_{\mathrm{orig}} / K_{\mathrm{MB}}$ for a $1.4\Msun$--$1.4\Msun$ BNS signal with several choices of data duration $T$ and a non-GR parameter value $\delta \hat{\varphi}_i$ used for calculating time to merger. The total frequency range is $20$--$2048\,\si{\hertz}$, and divided into frequency bands with $\{T^{(b)}\}_{b=0}^{B-1} = \{T,~T/2,~T/4,~\cdots,~4\,\si{\second}\}$.}
\begin{tabular}{c  c  c  c  c  c}
\hline \hline
\multirow{2}{*}{} &
\multirow{2}{*}{ $\delta \hat{\varphi}_{i}$} &
\multirow{2}{*}{$T~(\si{\second})$} &
\multirow{2}{*}{$K_{\mathrm{orig}}$} &
\multirow{2}{*}{$K_{\mathrm{MB}}$}&
\multirow{2}{*}{Speed up} \\ \\
\hline \hline
GR & $0$ & $256$ & $5.2\times10^5$ & $ 1.2\times 10^4$ & $4.5\times 10^1$  \\ 
\hline
\multirow{2}{*}{0PN ($i=0$)} & $20$ & $256$ & $5.2\times10^5$ & $2.8 \times 10^4$ & $1.8 \times 10^1$  \\
&  $20$ & $4096$ & $8.3\times10^6$ & $6.8\times 10^4$ & $1.2 \times 10^2$ \\
\hline
\multirow{2}{*}{-1PN ($i=-2$)}  &  $1$ & $256$ & $5.2\times10^5$ & $3.6 \times 10^4$ & $1.4 \times 10^1$  \\
& $1$ & $32768$ & $6.6\times10^7$ & $3.2\times10^5$ & $2.1\times10^2$  \\
\hline \hline
\end{tabular}
\label{tab:speed_up}
\end{table*}

\section{Validation} \label{sec:validation}

In this section, we study the accuracy of our technique using simulated BNS signals and real data.

\subsection{Simulation study} \label{sec:simulation}

\begin{table}
\captionsetup{labelfont=bf,
              justification=raggedright,
              singlelinecheck=false}
\caption{The injection values, prior, and explored range of GR parameters: Chirp mass $\mathcal{M}$, mass ratio $q \leq 1$, luminosity distance $D_L$, right ascension RA and declination DEC, orbital inclination angle $\theta_{JN}$, polarization angle $\psi$, constant phase $\phi_{\mathrm{c}}$, and coalescence time $t_{\mathrm{c}}$. The prior uniform in cosine, sine, and comoving volume are denoted by  ``Cosine", ``Sine" and ``Comoving" respectively. $t_{\mathrm{c,inj}}$ denotes the injection value of $t_{\mathrm{c}}$, and is set to the GPS time of $1187008882$ (17 Aug 2017, 12:41:04 UTC).}

\begin{ruledtabular}
\begin{tabular}{p{1.5cm} p{1cm} p{1.5 cm} p{1.5cm} p{1cm} p{1cm}}
Parameter & Unit &  Injection value & Prior & Min. & Max.\\
\hline
$\mathcal{M}$ & M$_\odot$ & 1.2 & Uniform & 1.15 & 1.25\\
$q$ & - & 0.8 & Uniform & 0.125 & 1\\
$\theta_{JN}$ &  rad. & 0.4 & Sine  & 0 & $\pi$\\
$D_L$ & Mpc & 72 & Comoving & $10$& $100$\\
RA & rad. & 3.45 &  Uniform & 0 & $2\pi$\\
DEC & rad. & $- 0.40$ & Cosine & $-\pi/2$ & $\pi/2$\\
$\psi$ & rad. & $0.65$ & Uniform & 0 & $\pi$\\
$\phi_{\mathrm{c}}$ & rad. & 1.3 & Uniform & 0 &$2\pi$\\
$t_{\mathrm{c}} - t_{\mathrm{c,inj}}$ & s & $0$ & Uniform & $- 0.1$ & $+ 0.1$
\end{tabular}
\end{ruledtabular}
\label{tab:bnspriors}
\end{table}

To verify that the multiband approximation does not bias the inference, we simulated BNS signals with non-zero $\delta \hat{\varphi}_i$, and performed parameterized tests on them with and without the multiband decomposition method.
The injection values, prior, and explored range of GR parameters are common among simulations, and outlined in \tabref{tab:bnspriors}.
The effects of spin angular momenta and tidal deformation of colliding objects were not taken into account for quick runs.
We considered the network of the two advanced LIGO detectors and the Virgo detector, and injected signals into Gaussian noise colored by their design sensitivities.
The analyzed frequency range is $20$--$2048\,\si{\hertz}$.
The network signal-to-noise ratios (SNRs) of the simulated signals are $\sim 50$.
The simulated signals were computed with the \texttt{TaylorF2} \cite{Buonanno:2009zt, Santamaria:2010yb} waveform model implemented in LAL, and the same waveform model was used for parameter recovery.

In this study, we considered two simulated signals: the $0\mathrm{PN}$ simulation with $\delta \hat{\varphi}_0=1,~\delta \hat{\varphi}_i=0~(i \neq 0)$ and the $-1\mathrm{PN}$ simulation with $\delta \hat{\varphi}_{-2}=0.003,~\delta \hat{\varphi}_i=0~(i \neq -2)$.
The duration of a signal from $20\,\si{\hertz}$ with vanishing non-GR parameters is $\sim 160\,\si{\second}$.
It is doubled for the $0\mathrm{PN}$ simulation or increased by $\sim 50\%$ for the $-1\mathrm{PN}$ simulation due to the non-zero GR parameter.
The explored parameter range is $-1 \leq \delta \hat{\varphi}_0 \leq 2$ for the $0$PN simulation and $-0.01 \leq \delta \hat{\varphi}_{-2} \leq 0.01$ for the $-1$PN simulation.
For each simulation, the prior of the non-GR parameter is uniform over its explored range.

The durations of analyzed data are $512\,\si{\second}$ and $256\,\si{\second}$ for the $0\mathrm{PN}$ and $-1\mathrm{PN}$ simulations respectively.
The total frequency range is divided into $8$ frequency bands with $\{T^{(b)}\}_{b=0}^7 = \{512\,\si{\second},~256\,\si{\second},~\cdots,~4\,\si{\second}\}$ for the multiband run of the $0\mathrm{PN}$ simulation, and $7$ frequency bands with $\{T^{(b)}\}_{b=0}^6 = \{256\,\si{\second},~128\,\si{\second},~\cdots,~4\,\si{\second}\}$ for the $-1\mathrm{PN}$ simulation.
The speed-up gains $K_{\mathrm{orig}} / K_{\mathrm{MB}}$ are $58$ and $37$ for the $0\mathrm{PN}$ and $-1\mathrm{PN}$ simulations respectively.

The stochastic sampling was performed with the \texttt{bilby} software and the \texttt{dynesty} \cite{Speagle:2019ivv} sampler.
The convergence of sampling is controlled by the number of live points $n_{\mathrm{live}}$ and the length of the MCMC chain in unit of its auto-correlation length $n_{\mathrm{ACT}}$ \cite{Romero-Shaw:2020owr}.
We used $n_{\mathrm{live}}=500, n_{\mathrm{ACT}}=10$ and $n_{\mathrm{live}}=1000, n_{\mathrm{ACT}}=10$ for the $0\mathrm{PN}$ and $-1\mathrm{PN}$ simulations respectively.
We have confirmed that increasing their values does not change the results significantly, which means the results are converged.
We marginalized the posterior over constant phase $\phi_{\mathrm{c}}$ analytically and luminosity distance $D_{\mathrm{L}}$ using the look-up table method \cite{Singer:2015ema, Thrane:2018qnx}.

\begin{figure}
                 \includegraphics[width = \columnwidth]{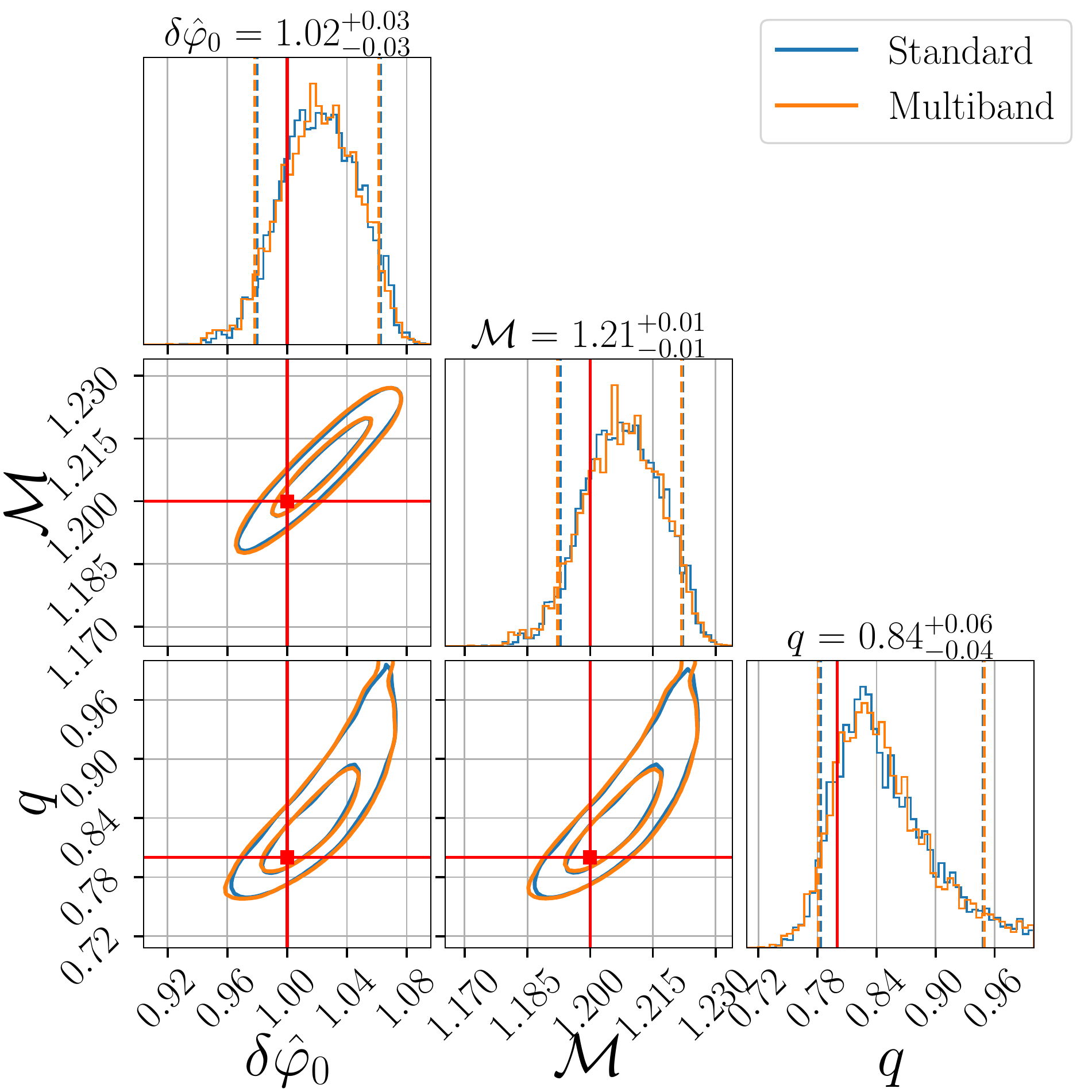}
                 \captionsetup{labelfont=bf,
              justification=raggedright,
              singlelinecheck=false}
                 \caption{One- and two-dimensional marginal posterior distributions of chirp mass $\mathcal{M}$, mass ratio $q$, and $0\mathrm{PN}$ relative deviation $\delta \hat \varphi_{0}$ from runs without (blue) and with (orange) the multiband decomposition technique. Diagonal panels show one-dimensional marginal posterior distributions, and vertical dashed lines indicate the $90\%$ credible intervals. Off-diagonal panels show two-dimensional marginal posterior distributions, and solid lines indicate the $50\%$ and $90\%$ credible regions. Red lines indicate the injection values.}
                 \label{fig:dchi0_few_lkl}
\end{figure}

\begin{figure}
                 \includegraphics[width = \columnwidth]{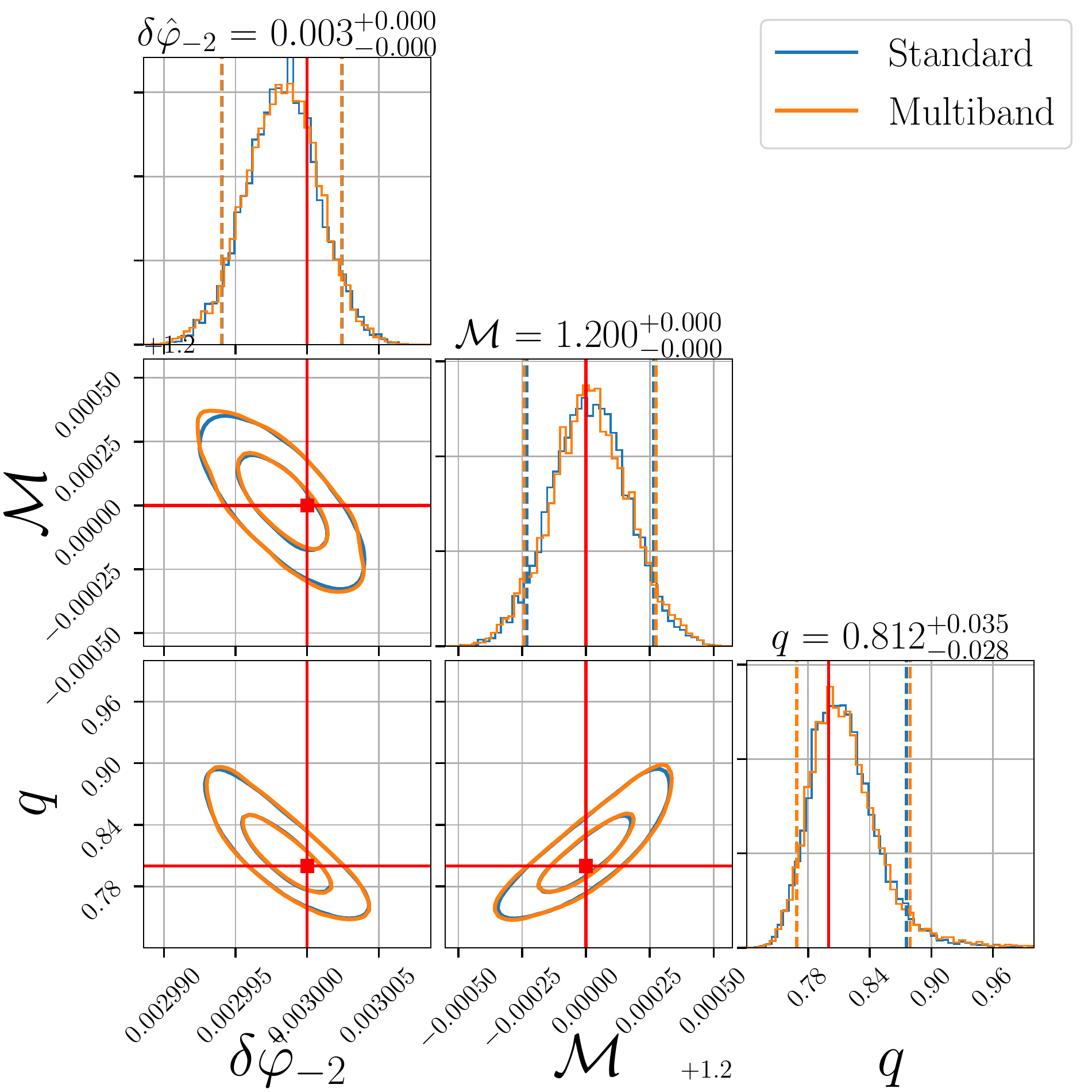}
                 \captionsetup{labelfont=bf,
              justification=raggedright,
              singlelinecheck=false}
                  \caption{One- and two-dimensional marginal posterior distributions of chirp mass $\mathcal{M}$, mass ratio $q$, and $-1\mathrm{PN}$ absolute deviation $\delta \hat \varphi_{-2}$ from runs without (blue) and with (orange) the multiband decomposition technique. Diagonal panels show one-dimensional marginal posterior distributions, and vertical dashed lines indicate the $90\%$ credible intervals. Off-diagonal panels show two-dimensional marginal posterior distributions, and solid lines indicate the $50\%$ and $90\%$ credible regions. Red lines indicate the injection values.}
                 \label{fig:dchimin2_few_lkl}
 \end{figure}

\begin{figure*}
                 \includegraphics[width = 7 in
                 ]{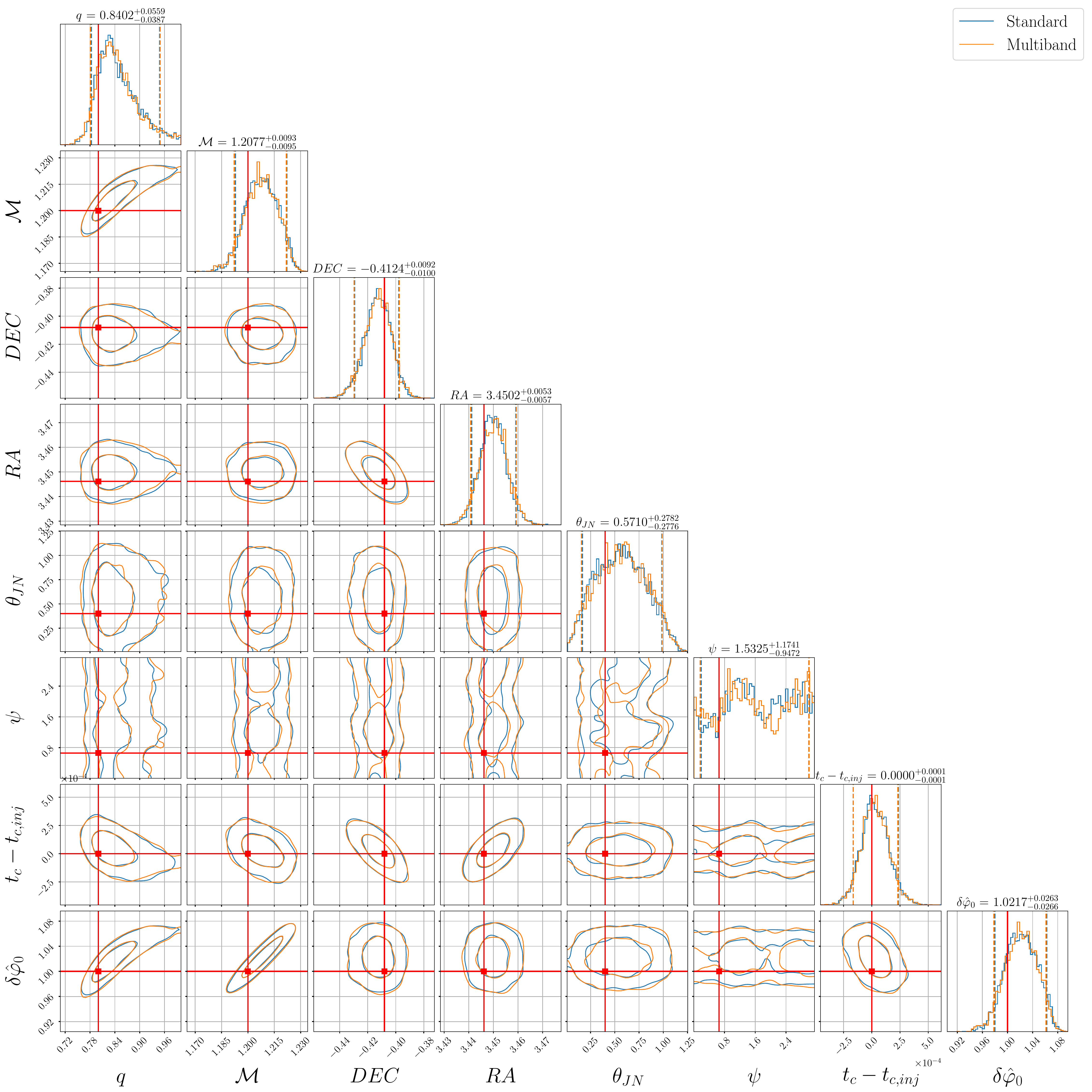}
                 \captionsetup{labelfont=bf,
              justification=raggedright,
              singlelinecheck=false}
                 \caption{One- and two-dimensional marginal posterior distributions of all the inferred parameters for the $0\mathrm{PN}$ simulation from runs without (blue) and with (orange) the multiband decomposition technique. Diagonal panels show one-dimensional marginal posterior distributions, and vertical dashed lines indicate the $90\%$ credible intervals. Off-diagonal panels show two-dimensional marginal posterior distributions, and solid lines indicate the $50\%$ and $90\%$ credible regions. Red lines indicate the injection values.}
                 \label{fig:likelihood_dchi0}
 \end{figure*}

\begin{figure*}
                 \includegraphics[width = 7.0
                 in ]{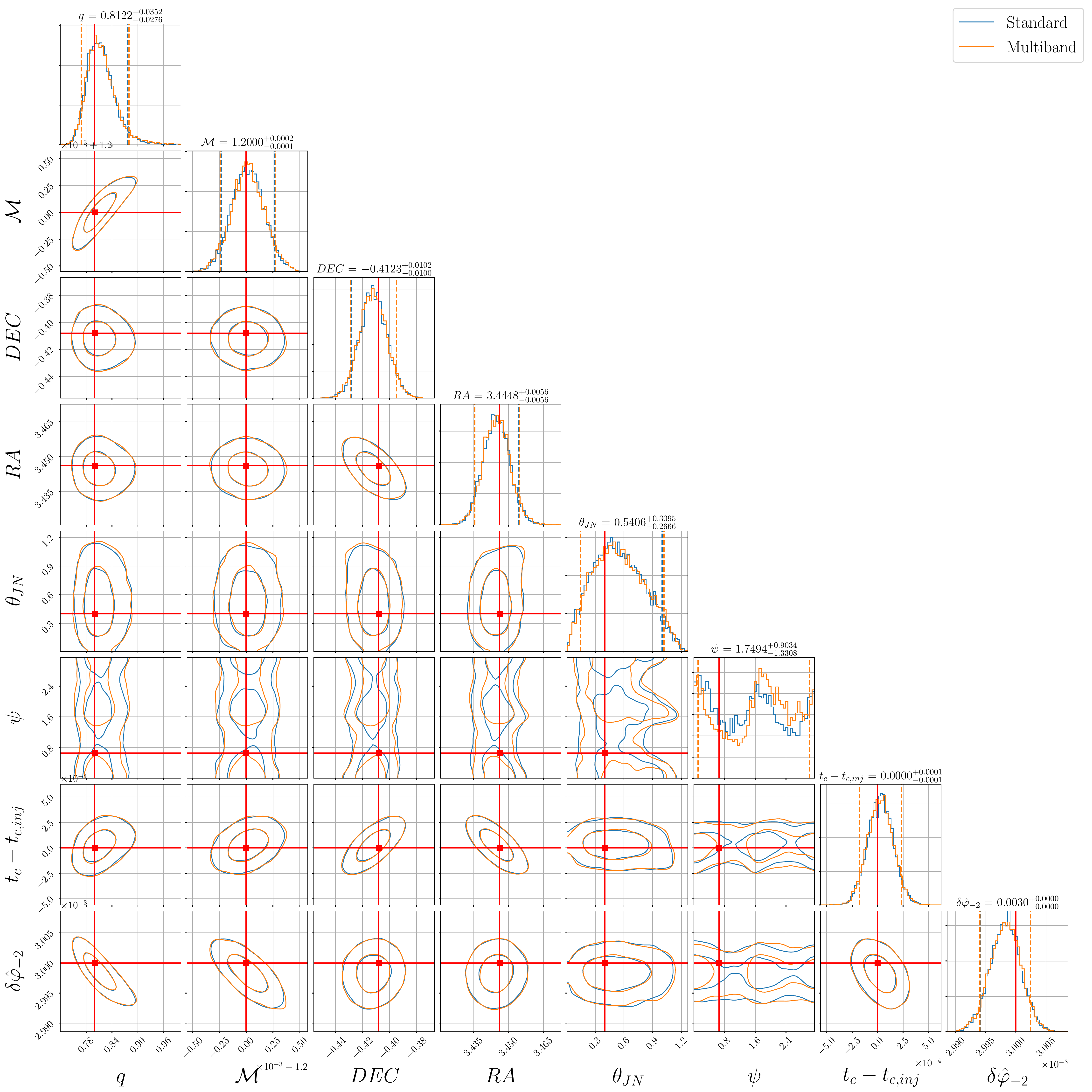}
                \captionsetup{labelfont=bf,
              justification=raggedright,
              singlelinecheck=false}
                 \caption{One- and two-dimensional marginal posterior distributions of all the inferred parameters for the $-1\mathrm{PN}$ simulation from runs without (blue) and with (orange) the multiband decomposition technique. Diagonal panels show one-dimensional marginal posterior distributions, and vertical dashed lines indicate the $90\%$ credible intervals. Off-diagonal panels show two-dimensional marginal posterior distributions, and solid lines indicate the $50\%$ and $90\%$ credible regions. Red lines indicate the injection values.}
                 \label{fig:likelihood_dchimin2}
 \end{figure*}

Figures \ref{fig:dchi0_few_lkl} and \ref{fig:dchimin2_few_lkl} show marginal posterior distributions of chirp mass ($\mathcal{M}$), mass ratio ($q$), and a non-GR parameter ($\delta \hat{\varphi}_0$ or $\delta \hat{\varphi}_{-2}$) for the $0\mathrm{PN}$ and $-1\mathrm{PN}$ simulations respectively.
The runs without and with the multiband approximation are labeled ``Standard" and ``Multiband" respectively.
As shown in the figures, the standard and multiband runs produce almost equivalent results in either simulation.
More quantitatively, the differences in the lower or upper bounds of the $90\%$ credible intervals are less than $4\%$ of their widths.
Those observations indicate that the multiband approximation is accurate enough for a relatively high SNR of $\sim 50$.
Since log-likelihood errors introduced by the multiband approximation are roughly proportional to the square of SNR, they are smaller for lower SNR values.
Therefore, our results show that our multiband approximation can be safely used also for SNR values below $50$.
The full posterior distributions of all the inferred parameters are presented in Figs \ref{fig:likelihood_dchi0} and \ref{fig:likelihood_dchimin2}.

The standard runs took $\sim 9$ days and $\sim 14$ days to complete for the $0\mathrm{PN}$ and $-1\mathrm{PN}$ simulations respectively.
They are reduced to $\sim 2$ hours and $\sim 7$ hours respectively with the multiband approximation.
The reduction of run times is more or less consistent with the speed-up gains estimated from $K_{\mathrm{orig}} / K_{\mathrm{MB}}$.
The runs were performed with an Intel Xeon Gold 6136 CPU with a clock rate of 3.0 GHz.
The stochastic sampling is parallelized with $48$ processes for the $0\mathrm{PN}$ simulation, and $24$ processes for the $-1\mathrm{PN}$ simulation.
 
\subsection{Likelihood errors for GW190814} \label{sec:gw190814}

For validating our approximation with more complicated signal morphology, we investigate the likelihood errors of our approximation for GW190814 \cite{LIGOScientific:2020zkf}.
We computed $\ln \mathcal{L}$ with and without our approximation on posterior samples from LIGO-Virgo parameter estimation analysis, and computed their differences $\Delta \ln \mathcal{L}$ as errors.
This signal is an appropriate test case for validating our approximation with gravitational-wave higher-order multiple moments since their effects are statistically significant for this signal \cite{LIGOScientific:2020zkf}.
We also considered the calibration uncertainties of detectors for validating our approximation with signal modulation caused by them.
The data were obtained from the Gravitational Wave Open Science Center \cite{gwosc}, and posterior samples were from \cite{tgrsamples}.
The \texttt{IMRPhenomPv3HM} \cite{Khan:2018fmp, Khan:2019kot} waveform model was employed for likelihood evaluations, which is the same model used for the LVK analysis.

\begin{figure*}[t]                                                              
        \begin{center}                                                          
                \includegraphics[width = 0.49\linewidth]{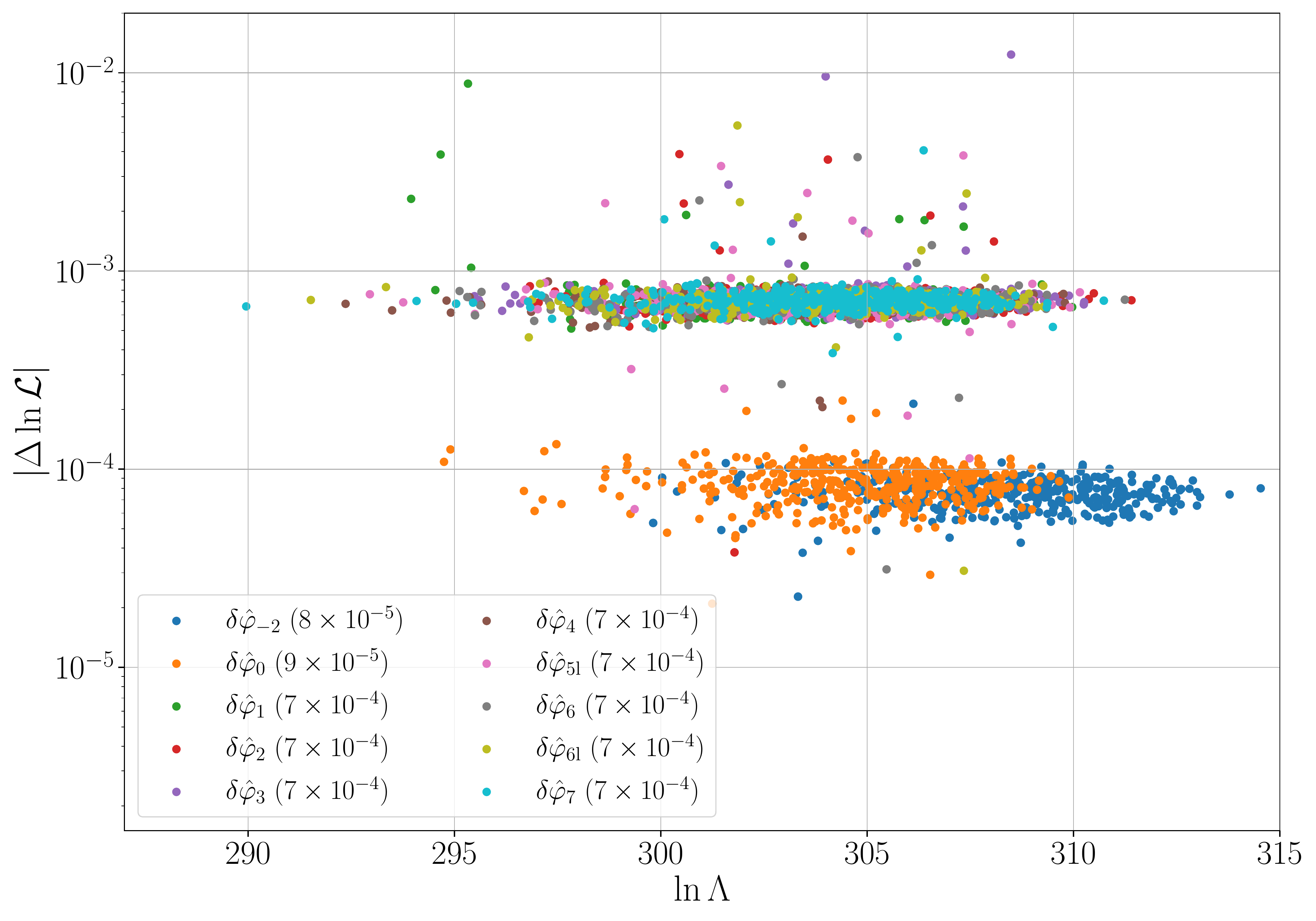}
                \includegraphics[width = 0.49\linewidth]{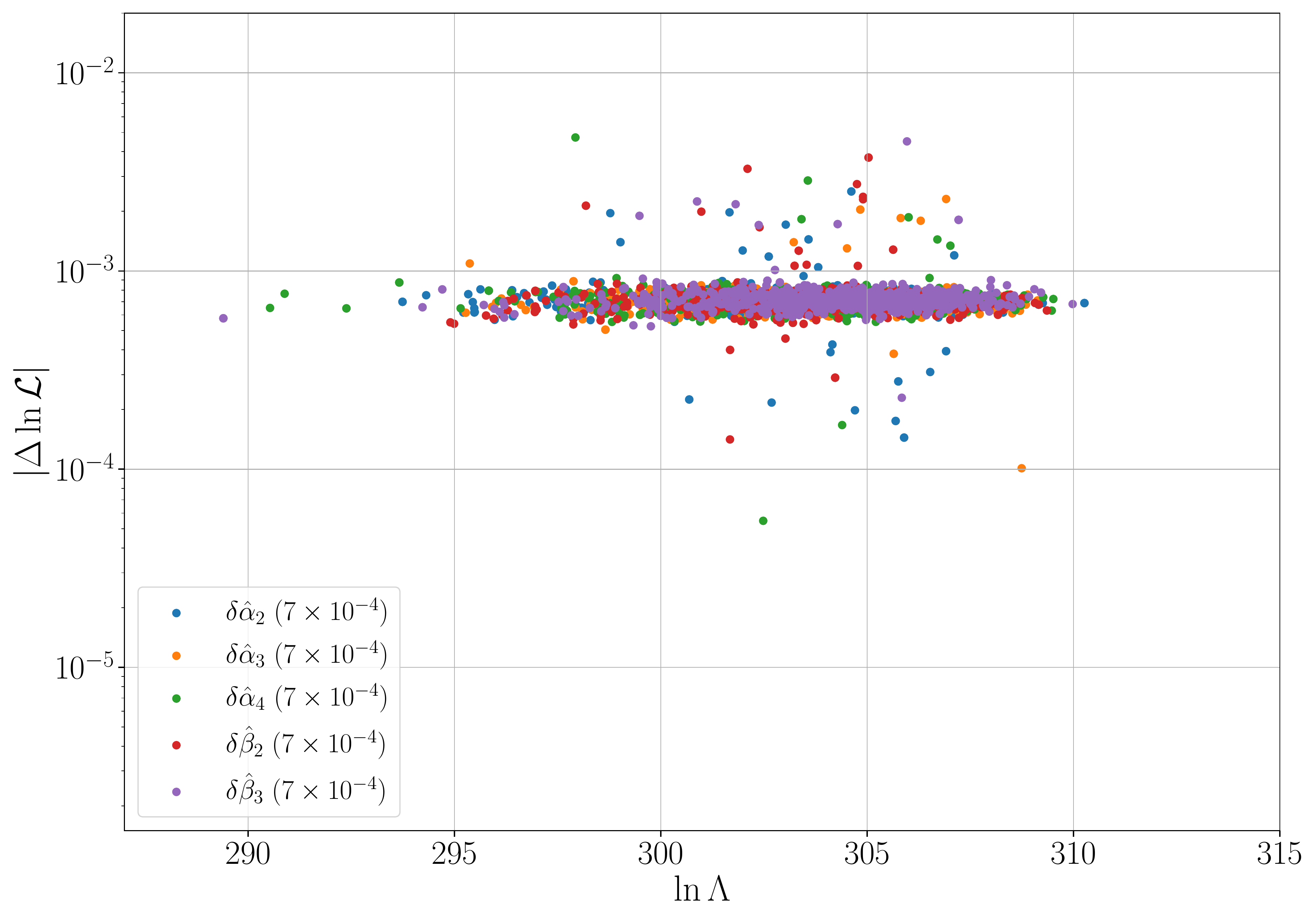}
                \captionsetup{labelfont=bf,
              justification=raggedright,
              singlelinecheck=false}
                \caption{Log-likelihood errors $|\Delta \ln \mathcal{L}|$ of the multiband approximation calculated on posterior samples from GW190814. The left plot shows the errors for tests of inspiral non-GR parameters and the right for tests of post-inspiral non-GR parameters. Each label shows the median value of these errors. }
                \label{fig:likelihood_errors}                                   
        \end{center}                                                            
\end{figure*}

Figure \ref{fig:likelihood_errors} shows $|\Delta \ln \mathcal{L}|$ with the horizontal axis representing the non-constant part of $\ln \mathcal{L}$,
\begin{equation}
\ln \Lambda \equiv \sum_i \left[\left(\bm{d}_i, \bm{h}_i\right)_i - \frac{1}{2} \left(\bm{h}_i, \bm{h}_i\right)_i\right].
\end{equation}
The left plot shows the errors for tests of inspiral parameters and the right for tests of post-merger parameters.
The total frequency range of $20$--$1024\,\si{\hertz}$ was divided into $3$ frequency bands with $\{T^{(b)}\}_{b=0}^2 = \{16\,\si{\second},~8\,\si{\second},~4\,\si{\second}\}$.
The frequency bands were determined based on the time-to-merger of the $m=4$ mode, and with the following reference values of $\mathcal{M}$, $\eta$, $\delta \hat \varphi_0$, and $\delta \hat \varphi_{-2}$,
\begin{equation}
\begin{aligned}
&\mathcal{M}=5.5M_{\odot},~~~\eta=0.25, \\
&\delta \hat \varphi_0 = \begin{cases}
20, & (\text{test of }\delta \hat \varphi_0) \\
0, & (\text{otherwise})
\end{cases} \\
&\delta \hat \varphi_{-2} = \begin{cases}
1, & (\text{test of }\delta \hat \varphi_{-2}) \\
0. & (\text{otherwise})
\end{cases}
\end{aligned}
\end{equation}
Those reference values were determined based on the parameter range explored by the LVK analysis.
The speed-up gain is $2.42$ for the test of $\delta \hat \varphi_0$, $2.44$ for the test of $\delta \hat \varphi_{-2}$, and $3.28$ for the other cases.
The {\it IFFT-FFT} algorithm was employed for computing $(\bm{h}, \bm{h})_i$ due to significant higher-order multiple moments.

Each plot label shows the median value of $|\Delta \ln \mathcal{L}|$.
The errors are $\lesssim 10^{-4}$ for the test of $\delta \hat \varphi_0$ or $\delta \hat \varphi_{-2}$, and $\lesssim 10^{-3}$ for the other tests.
The smaller errors for the former case are because frequency bands are constructed from a longer time-to-merger due to $\delta \hat \varphi_0>0$ or $\delta \hat \varphi_{-2}>0$.
In either case, the errors are much smaller than the unity, which shows our approximation is accurate enough for the analysis of GW190814.

\section{Conclusion} \label{sec:conclusion}

In this paper, we have presented a rapid inference technique for parameterized tests of GR, one of the main tests of GR using gravitational waves from CBC.
Our technique is based on a multiband decomposition of the gravitational-wave likelihood, which was originally developed for speeding up parameter estimation of CBC signals under the assumption of GR.
It exploits the chirping nature of a signal, and in principle is applicable to any chirp signal whose time to merger $\tau(f)$ is known.
To extend this technique to parameterized tests of GR, we have derived $\tau(f)$ taking into account non-GR deformations of the waveform.
Applying the multiband decomposition technique with our new formula of $\tau(f)$ to a $1.4\Msun$--$1.4\Msun$ BNS signal, we have found that our technique speeds up parameterized tests of a typical BNS signal by a factor of $\mathcal{O}(10)$ for the low-frequency cutoff of $20\,\si{\hertz}$. 

For validating our approximate technique, we have simulated BNS signals with SNRs of $\sim 50$.
Performing parameterized tests of them with and without our technique, we have verified that our technique produces results equivalent to those from runs without any approximate methods.
We have also computed log-likelihood errors of our technique for GW190814
and confirmed that they are well below unity.
Therefore, our work provides an efficient and accurate way of performing parameterized tests of GR, which is useful for dealing with more frequent detections in future observations.

We focus on single-parameter tests throughout this work. In principle, our technique can be applied to multiple-parameter tests using principal component analysis \cite{Shoom:2021mdj, Saleem:2021nsb}, with a modified time-to-merger formula parameterized by parameters corresponding to principal directions.
We leave that extension as a future work.

\begin{acknowledgments}
The authors thank an anonymous referee for helpful feedback. We thank Ignacio Maga\~{n}a Hernandez for reviewing the manuscript and providing valuable comments.
We also thank Jolien Creighton, Patrick Brady, and Brandon Piotrzkowski for their helpful comments on improving this paper.
The authors are supported by NSF PHY-2207728 and PHY-1912649.
The authors are grateful for computational resources provided by the Leonard E Parker Center for Gravitation, Cosmology and Astrophysics at the University of Wisconsin-Milwaukee and supported by NSF Grants PHY-1626190 and PHY-1700765.
The authors are grateful for computational resources provided by LIGO Laboratory and supported by NSF Grants PHY-0757058 and PHY-0823459.
This material is based upon work supported by NSF's LIGO Laboratory which is a major facility fully funded by the National Science Foundation.
This research has made use of data or software obtained from the Gravitational Wave Open Science Center (gw-openscience.org), a service of LIGO Laboratory, the LIGO Scientific Collaboration, the Virgo Collaboration, and KAGRA. LIGO Laboratory and Advanced LIGO are funded by the United States National Science Foundation (NSF) as well as the Science and Technology Facilities Council (STFC) of the United Kingdom, the Max-Planck-Society (MPS), and the State of Niedersachsen/Germany for support of the construction of Advanced LIGO and construction and operation of the GEO600 detector. Additional support for Advanced LIGO was provided by the Australian Research Council. Virgo is funded, through the European Gravitational Observatory (EGO), by the French Centre National de Recherche Scientifique (CNRS), the Italian Istituto Nazionale di Fisica Nucleare (INFN) and the Dutch Nikhef, with contributions by institutions from Belgium, Germany, Greece, Hungary, Ireland, Japan, Monaco, Poland, Portugal, Spain. The construction and operation of KAGRA are funded by Ministry of Education, Culture, Sports, Science and Technology (MEXT), and Japan Society for the Promotion of Science (JSPS), National Research Foundation (NRF) and Ministry of Science and ICT (MSIT) in Korea, Academia Sinica (AS) and the Ministry of Science and Technology (MoST) in Taiwan.
\end{acknowledgments}

\bibliographystyle{apsrev4-1}
\bibliography{reference}

 \end{document}